\documentclass[12pt,a4paper]{article}
\usepackage{amssymb}

\usepackage[left=5em]{geometry}
\usepackage{amsmath}
\usepackage{graphicx}

\title{Another Two Dark Energy Models Motivated from K$\acute{\text{a}}$rolyh$\acute{\text{a}}$zy
Uncertainty Relation}
\author{Cheng-Yi Sun\footnote{cysun@mailis.gucas.ac.cn; ddscy@163.com}\
$^{,1}$, Wen-Li Yang$^{1}$, Rui-Hong Yue$^{2}$ and Yu Song$^{1}$
\\
 {$^1$\small Institute of Modern Physics, Northwest University, Xian 710069, P.R.
 China.}\\
{$^2$\small Faculty of Science, Ningbo University, Ningbo 315211,
P.R. China.}}

\begin{document}
\maketitle
\begin{abstract}
The K$\acute{\text{a}}$rolyh$\acute{\text{a}}$zy uncertainty
relation indicates that there exists the minimal detectable cell
$\delta t^{3}$ over the region $t^3$ in Minkowski spacetime. Due to
the energy-time uncertainty relation, the energy of the cell $\delta
t^3$ can not be less $\delta t^{-1}$. Then we get a new energy
density of metric fluctuations of Minkowski spacetime as $\delta
t^{-4}$. Motivated by the energy density, we propose two new dark
energy models. One model is characterized by the age of the universe
and the other is characterized by the conformal age of the universe.
We  find that in the two models, the dark energy mimics a
cosmological constant in the late time.
\end{abstract}



\section{Introduction}

The study of dark energy has become one of the most active fields in
modern cosmology. Considerable efforts have been expended to explore
the nature of dark energy \cite{dark energy1,dark energy2}.
Recently, the agegraphic dark energy (ADE) \cite{ADE} and new
agegraphic dark energy (NADE) \cite{NADE} models are motivated  from
the K$\acute{\text{a}}$rolyh$\acute{\text{a}}$zy uncertainty
relation \cite{Karolyhzy,Ng1} which tells us that the time $t$ in
Minkowski space-time can not be known to a better accuracy than
\cite{Karolyhzy,Ng1} (see Ref.\cite{Ng2} for a recent review)
\begin{equation}
  \label{KUR}
  \delta t=\beta t_p^{2/3}t^{1/3}.
\end{equation}
Here $\beta$ is a dimensionless constant, $t_p$ is the reduced
Planck time. In this paper, we adopt the units $c=\hbar=1$.
Following \cite{0612110,0705.0924}, Eq.(\ref{KUR}) indicates that
for a length scale $t$, there exists a minimal detectable cell
$\delta t^3\sim t_p^2t$ over a region $t^3$. And the time-energy
uncertainty relation indicates that the energy of the minimal cell
can not be smaller than \cite{0612110,0705.0924,9903146}
\begin{equation}
  \label{ADEEUR}
  E_{\delta t^3}\sim t^{-1}.
\end{equation}
Then the energy density of the metric fluctuations of the Minkowski
space-time is \cite{0612110,0705.0924,9903146}
\begin{equation}
  \label{ADEMetricFluc}
    \rho_q\sim\frac{E_{\delta t^3}}{\delta t^3}\sim\frac{1}{t_p^2t^2}.
\end{equation}
Motivated by the equation, the energy density of ADE was proposed to
be \cite{ADE}
\[
  \rho_q=\frac{3n^2M_p^2}{T^2}.
\]
Here $n$ is a dimensionless constant parameter, $M_p=(8\pi
G)^{-1/2}$ and $T$ is the age of the universe
\begin{equation}
  \label{AgeOfUniv}
  T=\int^t_0{dt'}=\int^a_0{\frac{da}{Ha}},
\end{equation}
where $a$ is the scale factor, $H\equiv\dot{a}/a$ is the Hubble
parameter and a dot denotes the derivative with respect to the
cosmic time $t$. However, it is found that there exist some implicit
inconsistences in the model that ADE tracks the matter during the
matter-dominated epoch \cite{ADE,NADE} and the ability of ADE in
deriving the accelerated expansion contradicts the existence of the
radiation/matter-dominated epoch \cite{0708.2910,1008.0688}.

In order to address the drawbacks, the NADE model was proposed. In
NADE model, the energy density of dark energy was proposed to be
\cite{NADE}
\[
  \rho_q=\frac{3n^2M_p^2}{\eta^2},
\]
where $\eta$ is the conformal age of the universe
\begin{equation}
  \label{eta}
  \eta=\int^t_0{\frac{dt'}{a(t')}}=\int^a_0{\frac{da}{Ha^2}}.
\end{equation}
The NADE model is very successful in fitting the observation data
\cite{0904.0928,1011.6122}. However, in the NADE model, there is one
thing, which makes us uneasy, that the conformal age $\eta$ is not a
physical scale and can be rescaled arbitrarily.

We find that there may exist other ways out of the difficulties in
the ADE model. The relation (\ref{ADEMetricFluc}), which plays the
key role in the above motivations, is based on the
K$\acute{\text{a}}$rolyh$\acute{\text{a}}$zy uncertainty relation
(\ref{KUR}) and Eq.(\ref{ADEEUR}). The equation (\ref{ADEEUR}) is
motivated from the energy-time uncertainty relation. However, the
energy-time uncertainty relation may indicate the other natural
result. The K$\acute{\text{a}}$rolyh$\acute{\text{a}}$zy uncertainty
relation (\ref{KUR}) tells us that the time $t$ fluctuates with the
amplitude $\delta t\sim t_p^{2/3}t^{1/3}$. Due to the energy-time
uncertainty relation, this indicates the existence of energy
\begin{equation}
  \label{EDeltaT}
  E\sim\delta t^{-1}\sim\frac{1}{t_p^{2/3}t^{1/3}}.
\end{equation}
Actually, in \cite{0612110}, the above energy has been proposed and
supposed to distribut uniformly over the volume $t^3$. Then the
density energy $\rho_q\sim t_p^{-2/3}t^{-10/3}$ was derived
\cite{0612110}. Yet, this density energy cannot be used to motivate
a dark energy model since the corresponding fractional energy
density $\Omega_q\sim\rho_qt_p^{2}/H^{2}\sim t_p^{4/3}/t^{4/3}$ is a
decreasing function of $t$. However, if we suppose that the energy
given in Eq.(\ref{EDeltaT}) is distributed homogeneously over the
minimal detectable cell $\delta t^3\sim t_p^{2}t$, then we get the
energy density $\rho_q$ associated with the fluctuation of the
Minkowski space-time as
\begin{equation}
  \label{AADEMetricFluc}
  \rho_q\sim\frac{E}{\delta t^3}\sim\frac{1}{\delta
  t^4}\sim\frac{1}{t_p^{8/3}t^{4/3}}.
\end{equation}
Fortunately, we find that from the above energy density, new dark
energy models can be motivated.

Based on Eq.(\ref{AADEMetricFluc}), in Sec.\ref{SecAADE}, we propose
a new dark energy model characterized by the age of the universe to
address the drawbacks in the ADE model. In Sec.\ref{SecANADE},
following the motivation of NADE, we motivate a new dark energy
model characterized by the conformal age of the universe from
Eq.(\ref{AADEMetricFluc}). Finally, the summary is given.

\section{New Model Characterized by Age of the Universe}
\label{SecAADE}

Based on Eq.(\ref{AADEMetricFluc}),  following the motivation of the
ADE model we may propose a new model of dark energy characterized by
the age of the universe as
\begin{equation}
  \label{AADE}
  \rho_q=\frac{3n^{2/3}M_p^{8/3}}{T^{4/3}},
\end{equation}
where $n$ is a constant parameter and $T$ is the age of the universe
defined in Eq.(\ref{AgeOfUniv}). The corresponding fractional energy
density is
\begin{equation}
  \label{FracED}
  \Omega_q\equiv\frac{n^{2/3}M_p^{2/3}}{T^{4/3}H^2}.
\end{equation}
Let us show whether the dark energy (\ref{AADE}) can drive the
present accelerated expansion of the universe or not. In the
matter-dominated (MD) epoch, since $H=\frac{2}{3t}\propto a^{-3/2}$,
then we have $\Omega_q\propto T^{2/3}\propto a$. Then the fractional
energy density (\ref{FracED}) increases during the MD epoch and the
dark energy (\ref{AADE}) becomes dominated eventually. So the
tracking behavior of ADE during the MD epoch does not exist in this
model. From Eq.(\ref{AADE}), we have
\begin{equation}
  \label{dAADEdt}
  \dot{\rho}_q+\frac{4}{3}\frac{\rho_q}{T}=0.
\end{equation}
By comparing the equation with the conservation law of the dark
energy $\dot{\rho}_q+3H(1+w_q)\rho_q=0$ and using Eq.(\ref{FracED}),
we can get the equation of state (EoS) parameter $w_q$ as
\begin{equation}
  \label{EoSq}
  w_q=-1+\frac{4}{9}\Omega_q^{3/4}\sqrt{\frac{H}{nM_p}}.
\end{equation}
We can derive from Eq.(\ref{EoSq}) that in the dark-energy-dominated
(DED) epoch the parameter $w_q$ is given by
\begin{equation}
  \label{EoSDM}
  w_q\simeq-1+\frac{4}{9}\frac{1}{(nM_pT)^{1/3}},
\end{equation}
since approximately $\Omega_q\simeq1$ and
$H^2\simeq\frac{n^{2/3}M_p^{2/3}}{T^{4/3}}$. Then, as the increasing
of $T$, $w_q$ approaches $-1$ and the dark energy (\ref{AADE})
mimics a cosmological constant to drive the accelerated expansion.
So, from the above analysis, we know that the dark energy
(\ref{AADE}) has at least the reasonable qualitative behavior.

\begin{figure}
\centering
\renewcommand{\figurename}{Fig.}
\includegraphics[scale=0.85]{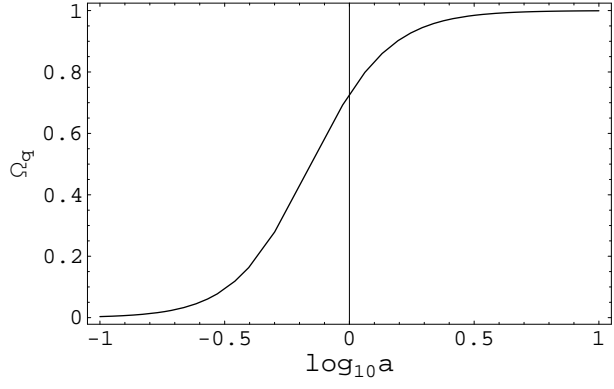}
\caption{The evolution of $\Omega_q$ for the dark energy
(\ref{AADE}) versus $\log_{10}{a}$ with the initial condition
$\Omega_{q0}=0.728$ and $\tilde{n}=20$.\label{FigAADEOmegaq}}
\end{figure}

Now considering the flat Friedmann-Robertson-Walk (FRW) universe
with the dark energy (\ref{AADE}) and pressureless matter, the
corresponding Friedmann equation reads
\begin{equation}
  \label{FE}
  H^2=\frac{1}{3M_p^2}(\rho_m+\rho_q).
\end{equation}
Here $\rho_m$ is the energy density of the matter and the
corresponding conservation law is
\begin{equation}
  \label{CLMatter}
  \dot{\rho}_m+3H\rho_m=0.
\end{equation}
Then using Eqs.(\ref{FracED}), (\ref{FE}), (\ref{AADE}) and
(\ref{CLMatter}), we find that the evolution of the fractional
energy density (\ref{FracED}) is governed by the two equations
\begin{align}
  \label{dOmegaqDaTildeH}
  \frac{d\Omega_q}{da}&=\frac{3}{a}\Omega_q(1-\Omega_q)\Big(1-\frac{4}{9}\Omega_q^{3/4}\sqrt{\frac{\tilde{H}}{\tilde{n}}}\ \Big),\\
  \label{dTHda}
  \frac{d\tilde{H}}{da}&=-\frac{3\tilde{H}}{2a}\Big(1-\Omega_q+\frac{4}{9}\Omega_q^{7/4}\sqrt{\frac{\tilde{H}}{\tilde{n}}}\ \Big),
\end{align}
where
\begin{equation}
  \label{tildeHn}
  \tilde{H}\equiv\frac{H}{H_0},\quad
  \tilde{n}\equiv\frac{nM_p}{H_0},
\end{equation}
and the subscript $0$ denotes the present value of the corresponding
parameter. By choosing $a_0=1$, $\tilde{n}=20$  and the initial
condition $\Omega_{q0}=0.728$, we solve the equations and display
the evolution of $\Omega_q$ with respect to $\log_{10}{a}$ in
Fig.\ref{FigAADEOmegaq}. The result displayed in
Fig.\ref{FigAADEOmegaq} tells us that the dark energy (\ref{AADE})
is negligible in the early universe and becomes dominated in the
late, which is consistent with the analysis in the last paragraph.

\begin{figure}
\centering
\renewcommand{\figurename}{Fig.}
\includegraphics[scale=0.85]{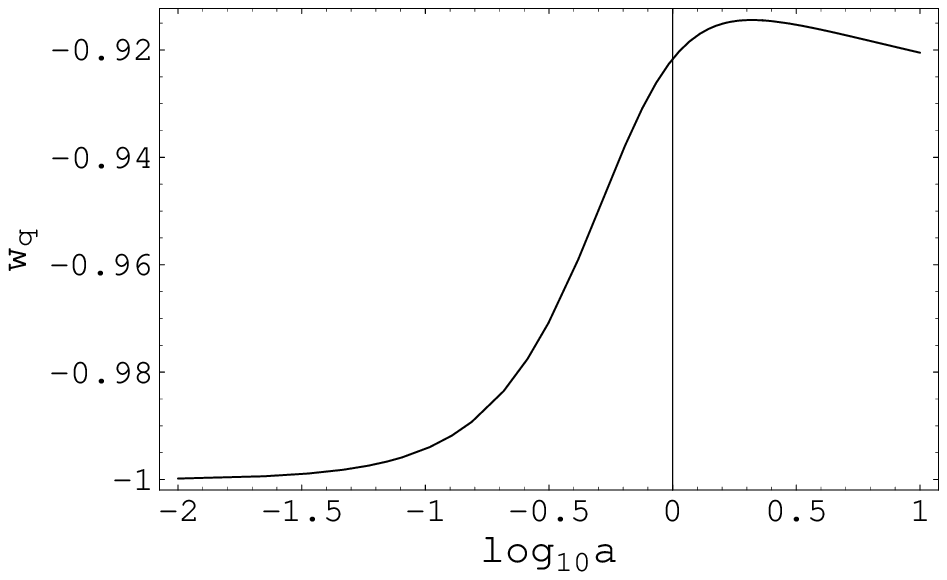}
\caption{The evolution of $w_q$ for the dark energy (\ref{AADE})
versus $\log_{10}{a}$ with the initial condition $\Omega_{q0}=0.728$
and $\tilde{n}=20$.\label{FigAADEwa}}
\end{figure}

\begin{figure}
\centering
\renewcommand{\figurename}{Fig.}
\includegraphics[scale=0.75]{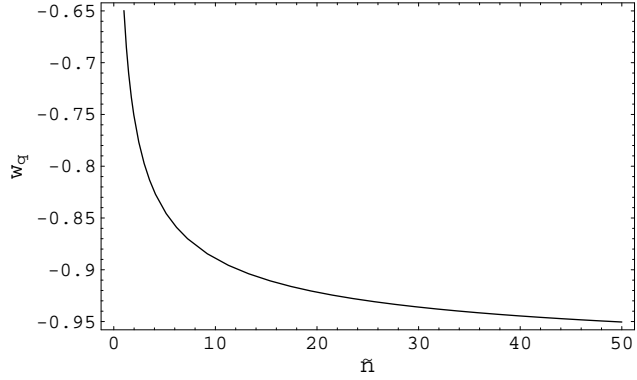}
\caption{The current equation of state parameter $w_q$ versus the
parameter $\tilde{n}$ with fixed
$\Omega_{q0}=0.728$.\label{FigAADEwn}}
\end{figure}

Using Eq.(\ref{tildeHn}), we may rewrite Eq.(\ref{EoSq}) as
\begin{equation}
  \label{EoSqTildeHn}
  w_q=-1+\frac{4}{9}\Omega_q^{3/4}\sqrt{\frac{\tilde{H}}{\tilde{n}}}\ .
\end{equation}
Then using the above equation and the numerical solution obtained by
solving Eqs.(\ref{dOmegaqDaTildeH}) and (\ref{dTHda}) with $a_0=1$,
$\tilde{n}=20$  and $\Omega_{q0}=0.728$, we display the evolution of
$w_q$ with respect to $\log_{10}{a}$ in Fig.\ref{FigAADEwa}. We also
plot the current value of $w_{q}$ versus $\tilde{n}$ with fixed
$\Omega_{q0}=0.728$ in Fig.\ref{FigAADEwn}. We see that
$w_{q0}\lesssim -0.89$ as $\tilde{n}\gtrsim 10$. Therefore the EoS
parameter is consistent with the Wilkinson Microwave Anisotropy
Probe(WMAP) observation \cite{WMAP7Year}, as $\tilde{n}$ is taken to
be a number of order ten.

Furthermore, let us calculate the shift parameter $R$, which
characterizes the position of the first peak of the cosmic microwave
background spectrum and is defined as \cite{shiftPara}
\begin{equation}
  \label{shiftParas}
  R=\sqrt{\Omega_{m0}}\int^{z_*}_0{\frac{dz}{\tilde{H}(z)}}.
\end{equation}
Here $z_*$ is the redshift of decoupling. The 7-year WMAP
observations tell us that $z_*=1091.3\pm0.91$ at $1\sigma$
confidence level \cite{WMAP7Year}. In this paper, we choose
$z_*=1091$. Then, using the numerical solution of
Eqs.(\ref{dOmegaqDaTildeH}) and (\ref{dTHda}) with $a_0=1$,
$\tilde{n}=20$  and $\Omega_{q0}=0.728$, we obtain the value of $R$
to be
\begin{equation}
  \label{AADEShiftPara}
  R=1.722.
\end{equation}
The 7-year WMAP observations tell us $R=1.725\pm0.018$ at $1\sigma$
confidence level \cite{WMAP7Year}. So the dark energy model
(\ref{AADE}) with $\tilde{n}=20$ is consistent with the 7-year WMAP
observations.

\begin{figure}
\centering
\renewcommand{\figurename}{Fig.}
\includegraphics[scale=0.8]{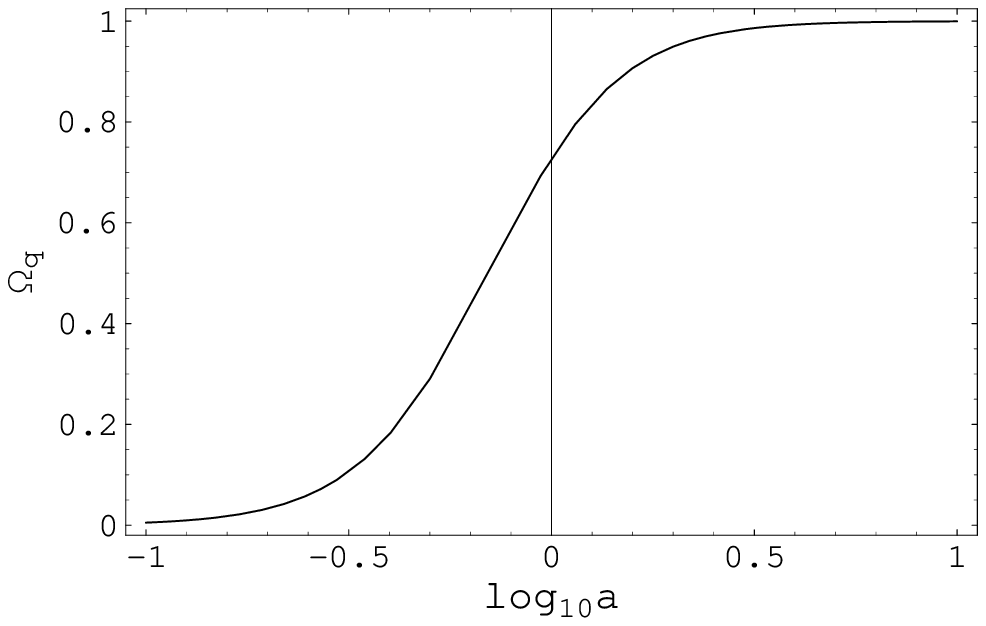}
\caption{The evolution of $\Omega_q$ for the dark energy
(\ref{ANADE}) versus $\log_{10}{a}$ with the initial condition
$\Omega_{q0}=0.728$ and $\tilde{n}=20$.\label{FigANADEOmegaq}}
\end{figure}

\begin{figure}
\centering
\renewcommand{\figurename}{Fig.}
\includegraphics[scale=0.85]{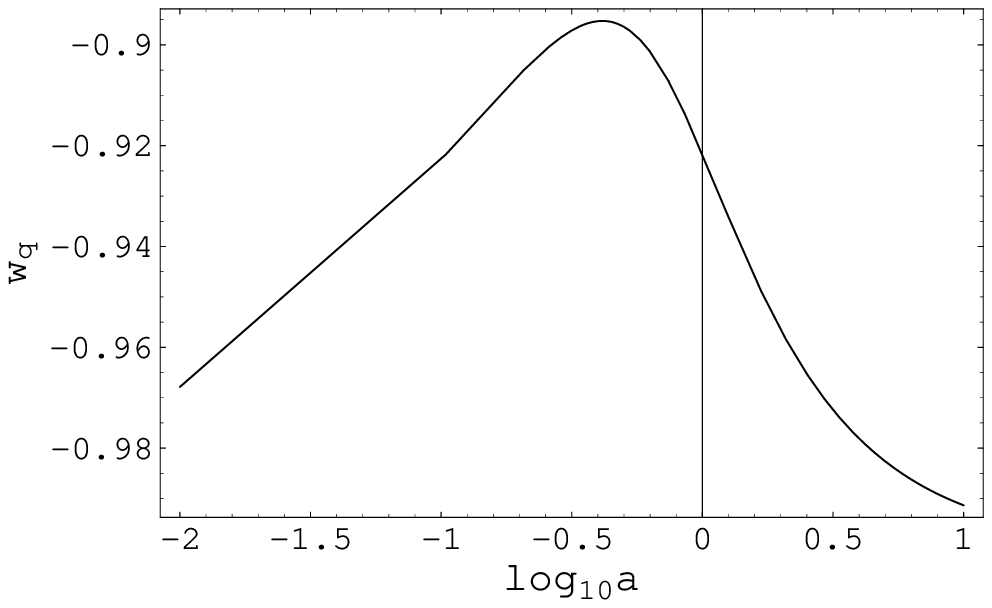}
\caption{The evolution of $w_q$ for the dark energy (\ref{ANADE})
versus $\log_{10}{a}$ with the initial condition $\Omega_{q0}=0.728$
and $\tilde{n}=20$.\label{FigANADEwa}}
\end{figure}

\section{New Model Characterized by Conformal Age of Universe}
\label{SecANADE}

In order to eliminate the inconsistences in the ADE model, the
authors in Ref.\cite{NADE} proposed the NADE model by replacing the
age $T$ with the conformal age $\eta$. The NADE model is very
successful in fitting the observation data
\cite{0904.0928,1011.6122}. Motivated by this, from
Eq.(\ref{AADEMetricFluc}) we propose a dark energy model
characterized by the conformal age of the universe as
\begin{equation}
  \label{ANADE}
  \rho_q=\frac{3n^{2/3}M_p^{8/3}}{\eta^{4/3}},
\end{equation}
where $\eta$ is the conformal age of universe defined in
Eq.(\ref{eta}). The corresponding fractional energy density is
\begin{equation}
  \label{FracEDANADE}
  \Omega_q=\frac{n^{2/3}M_p^{2/3}}{\eta^{4/3}H^2}.
\end{equation}

Let us consider the flat FRW universe filled with the dark energy
(\ref{ANADE}) and the pressureless matter. In the MD epoch, the
energy conservation equation of the matter, Eq.(\ref{CLMatter}),
tells us $H^2\propto a^{-3}$. Substituting this into Eq.(\ref{eta}),
we have $\eta\propto\sqrt{a}$. Then Eq.(\ref{FracEDANADE}) tells us
that in the MD epoch, $\Omega_q\propto a^{7/3}$. Then we know that
the fractional energy density (\ref{FracEDANADE}) increases in the
matter-dominated epoch and eventually, the dark energy (\ref{ANADE})
becomes dominated. From Eqs.(\ref{ANADE}), (\ref{eta}),
(\ref{FracEDANADE}) and the conservation equation
$\dot{\rho}_q+3H(1+w_q)\rho_q=0$, we can easily get the EoS
parameter as
\begin{equation}
  \label{EoSANADE}
  w_q=-1+\frac{4}{9}\frac{\Omega_q^{3/4}}{a}\sqrt{\frac{H}{nM_p}}.
\end{equation}
In the DED epoch, since approximately $\Omega_q\simeq1$ and
$H^2\simeq\frac{n^{2/3}M_p^{2/3}}{\eta^{4/3}}$, from
Eq.(\ref{EoSANADE}) we have
\begin{equation}
  \label{EoSANADEDD}
  w_q\simeq-1+\frac{4}{9}\frac{1}{a(nM_p\eta)^{1/3}}.
\end{equation}
So, as the expansion of the universe, $w_q$ approaches $-1$ and the
dark energy (\ref{ANADE}) mimics a cosmological constant. The above
analysis makes us believe that qualitatively the dark energy
(\ref{ANADE}) is a reasonable model.

In order to confirm the qualitative analysis in the last paragraph,
now let us survey the evolution of the fractional energy density
(\ref{FracEDANADE}) quantitatively. Using Eqs.(\ref{eta}),
(\ref{FE}), (\ref{ANADE}), (\ref{FracEDANADE}) and (\ref{CLMatter}),
we find that the evolution of $\Omega_q$ is governed by the two
equations
\begin{align}
  \label{dOmegaqANADEDaTildeH}
  \frac{d\Omega_q}{da}&=\frac{3}{a}\Omega_q(1-\Omega_q)\Big[1-\frac{4}{9}\frac{\Omega_q^{3/4}}{a}\sqrt{\frac{\tilde{H}}{\tilde{n}}}\ \Big],\\
  \label{dTHetaDa}
  \frac{d\tilde{H}}{da}&=-\frac{3\tilde{H}}{2a}\Big[1-\Omega_q+\frac{4}{9}\frac{\Omega_q^{7/4}}{a}\sqrt{\frac{\tilde{H}}{\tilde{n}}}\
  \Big],
\end{align}
where $\tilde{H}\equiv H/H_0$ and $\tilde{n}\equiv nM_p/H_0$. Still
choosing $a_0=1$, $\tilde{n}=20$ and $\Omega_{q0}=0.728$, we solve
the equations numerically and display the evolution of $\Omega_q$
with respect to $\log_{10}{a}$ in Fig.\ref{FigANADEOmegaq}. The
result displayed in Fig.\ref{FigANADEOmegaq} tells us that the dark
energy (\ref{ANADE}) is negligible in the early universe and becomes
dominated in the late, which confirms the qualitative analysis in
the last paragraph.

By rewriting Eq.(\ref{EoSANADE}) as
\begin{equation}
  \label{EoSANADETildelHn}
  w_q=-1+\frac{4}{9}\frac{\Omega_q^{3/4}}{a}\sqrt{\frac{\tilde{H}}{\tilde{n}}},
\end{equation}
we display the evolution of $w_q$ versus $\log_{10}{a}$ in
Fig.\ref{FigANADEwa} with $a_0=1$, $\tilde{n}=20$ and
$\Omega_{q0}=0.728$. From Fig.\ref{FigANADEwa}, we know that $w_q$
approaches $-1$ in the future, which confirms the qualitative
analysis. With $a_0=1$, we find that the curve representing the
current value of $w_q$ defined in Eq.(\ref{EoSANADETildelHn}) versus
$\tilde{n}$ with $\Omega_{q0}=0.728$ is just that displayed in
Fig.\ref{FigAADEwn}. So, as in the last section,  from
Fig.\ref{FigAADEwn} we may conclude that the EoS parameter of the
dark energy (\ref{ANADE}) is consistent with the WMAP observation
\cite{WMAP7Year}, as $\tilde{n}$ is taken to be a number of order
ten. Further, with $a_0=1$, $\tilde{n}=20$ and $\Omega_{q0}=0.728$,
we obtain the value of the shift parameter $R$ as
\begin{equation}
  \label{ANADEShiftPara}
  R=1.716.
\end{equation}
So the dark energy model (\ref{ANADE}) with $\tilde{n}=20$ is in
agreement with the 7-year WMAP observations ($R=1.725\pm0.018$)
\cite{WMAP7Year}.

\section{Summary}

We propose two new models of dark energy based on the
K$\acute{\text{a}}$rolyh$\acute{\text{a}}$zy uncertainty relation
(\ref{KUR}). The two models are different from the ADE \cite{ADE}
and NADE \cite{NADE} models. The ADE and NADE models are motivated
from Eq.(\ref{ADEMetricFluc}) which is deduced from Eq.(\ref{KUR})
by taking the energy of the minimal detectable cell $\delta t^3$ to
be $t^{-1}$, while the two models in this note are motivated from
Eq.(\ref{AADEMetricFluc}) which is obtained by arguing the energy of
the cell $\delta t^3$ to be $\delta t^{-1}$. Both
Eq.(\ref{ADEMetricFluc}) and Eq.(\ref{AADEMetricFluc}) are the
natural results of the relation (\ref{KUR}) and the energy-time
uncertainty relation. So, following the motivation of the ADE and
NADE models, we motivate the two models from
Eq.(\ref{AADEMetricFluc}).

It is well known that there exist implicit inconsistences in the
model of ADE \cite{ADE,NADE,0708.2910,1008.0688}. In fact, it is in
order to eliminate the inconsistences of ADE that the model of NADE
is proposed \cite{NADE}.  We find that in the two models in the
note, the dark energy has the reasonable behavior that the dark
energy is negligible in the MD epoch and eventually becomes
dominated to derive the accelerated expansion, and no inconsistences
exist in the two model. Particularly, the model proposed in
Sec.\ref{SecAADE} is characterized by the age of the universe $T$,
not the unphysical scale $\eta$.

By calculating the shift parameter, we compare the two models with
the observational data. We find the two models with $\tilde{n}=20$
fit the 7-year WMAP date well. We hope that the two models can shed
new light on the solving of the dark energy problem.

\section*{Acknowledge}
This work has been supported in part by the NNSF of China under
Grant No.11147017, the Research Fund for the Doctoral Program of
Higher Education of China under Grant No.20106101120023, the NNSF of
China under Grant No.10875060 and the Natural Science Foundation of
the Shaanxi Province under Grand No.2011JQ1002.

\end{document}